\documentclass[10pt,A4paper,conference]{IEEEtran}
\begin{document}

\title{Secret Key and Private Key Constructions for Simple Multiterminal Source Models}

\author{\authorblockN{Chunxuan Ye}
\authorblockA{Department of Electrical and Computer Engineering\\
and Institute for Systems Research\\
University of Maryland\\
College Park, MD 20742, USA \\
E-mail: cxye@eng.umd.edu}
\and
\authorblockN{Prakash Narayan}
\authorblockA{Department of Electrical and Computer Engineering\\
and Institute for Systems Research\\
University of Maryland\\
College Park, MD 20742, USA \\
E-mail: prakash@eng.umd.edu} 
 }

\maketitle

\begin{abstract}
This work is motivated by recent results of Csisz\'ar and Narayan 
({\it IEEE Trans. on Inform. Theory, Dec. 2004}), which highlight 
innate connections between secrecy generation by multiple terminals 
and multiterminal Slepian-Wolf near-lossless data compression (sans 
secrecy restrictions). We propose a new approach for {\it constructing} 
secret and private keys based on the long-known Slepian-Wolf code for 
sources connected by a virtual additive noise channel, due to Wyner 
({\it IEEE Trans. on Inform. Theory, Jan. 1974}). Explicit procedures 
for such constructions, and their substantiation, are provided.
\end{abstract}

\section{Introduction}
The problem of secret key generation by multiple terminals, based on 
their observations of distinct correlated signals followed by public 
communication among themselves, has been investigated by several 
authors (\cite{Mau93, AhlCsi93}, among others). It has been shown 
that these terminals can generate common randomness which is kept 
secret from an eavesdropper privy to the public interterminal 
communication. Of particular relevance to us are recent results in 
\cite{CsiNar04} for models with an arbitrary number of terminals, 
each of which observes a distinct component of a discrete memoryless 
multiple source (DMMS). Unrestricted public communication is allowed 
between these terminals. All the transmissions are observed by all 
the terminals and by the eavesdropper. Two models considered in 
\cite{CsiNar04} are directly relevant to our work, and these are 
first briefly described below.

\noindent (i) Suppose that $d\geq 2$ terminals observe $n$ i.i.d. 
repetitions of the random variables (rvs) $X_1, \cdots X_d$, denoted 
by ${\bf X}_1, \cdots, {\bf X}_d$, respectively. A secret key (SK) 
generated by these terminals consists of ``common randomness,'' 
based on {\it public} interterminal communication, which is concealed 
from an eavesdropper with access to this communication. The largest 
(entropy) rate of such a SK is termed the SK-capacity, denoted by 
$C_{SK}$, and is shown in \cite{CsiNar04} to equal
\begin{equation}
C_{SK}=H(X_1, \cdots ,X_d)-R_{min},
\label{e1}
\end{equation}
where 
\[
R_{min}=\min_{(R_1, \cdots , R_d)\in {\cal R}}\sum_{i=1}^dR_i,
\]
with 
\begin{eqnarray*}
{\cal R}&=&\{(R_1, \cdots , R_d): \sum_{i\in B} R_i\geq \\
      && H(\{X_j,\ j\in B\}|\{X_j,\  j\in B^c\}), B\subset \{1, \cdots, d\}\},
\end{eqnarray*}
where $B^c=\{1,\cdots ,d\}\backslash B$.

\noindent (ii) For a given subset $A\subset \{1, \cdots , d\}$, a 
private key (PK) for the terminals in $A$, private from the terminals 
in $A^c$, is a SK generated by the terminals in $A$ (with the possible 
help of the terminals in $A^c$), which is concealed from an eavesdropper 
with access to the public interterminal communication and also from the 
``helper'' terminals in $A^c$ (and, hence, private). The largest (entropy) 
rate of such a PK is termed the PK-capacity, denoted by $C_{PK}(A)$. It 
is shown in \cite{CsiNar04} that 
\begin{equation}
C_{PK}(A)=H(\{X_i,\ i\in A\}|\{X_i,\ i\in A^c\})-R_{min}(A),
\label{e2}
\end{equation}
where 
\[
R_{min}(A)=\min_{\{R_i,i\in A\}\in {\cal R}(A)}\sum_{i\in A} R_i,
\]
with 
\begin{eqnarray*}
{\cal R}(A)&=&\{ \{R_i, i\in A\}: \sum_{i\in B} R_i\geq \\
           &&H(\{X_j,\ j\in B\}|\{X_j,\  j\in B^c\}), B\subset A\}.
\end{eqnarray*}

The results above afford the following interpretation. The SK-capacity 
$C_{SK}$, i.e., largest rate at which all the $d$ terminals can generate 
a SK, is obtained by subtracting from the maximum rate of shared common 
randomness achievable by these terminals, viz. $H(X_1, \cdots , X_d)$, 
the smallest sum-rate $R_{min}$ of the data-compressed interterminal 
communication which enables each of the terminals to acquire this maximal 
common randomness. A similar interpretation holds for the PK-capacity 
$C_{PK}(A)$ as well, with the difference that the terminals in $A^c$, 
which act as helpers but must not be privy to the secrecy generated, 
can simply ``reveal'' their observations. Hence, the entropy terms in 
(\ref{e1}) are now replaced in (\ref{e2}) with additional conditioning 
on $\{X_i,\ i\in A^c\}$. It should be noted that $R_{min}$ and $R_{min}(A)$ 
are obtained as solutions to Slepian-Wolf (SW) multiterminal near-lossless 
data compression problems {\it not involving any secrecy constraints}. 
This characterization of the SK-capacity and PK-capacity in terms of the 
decompositions above also mirrors the consecutive stages in the random 
coding arguments used in establishing these results. For instance, and 
loosely speaking, to generate a SK, the $d$ terminals first generate 
common randomness (without any secrecy restrictions), say a rv $L$ of 
entropy rate $\frac{1}{n}H(L)>0$, through SW-compressed interterminal 
communication ${\bf F}$. This means that all the $d$ terminals acquire 
the rv $L$ with probability $\cong 1$. The next step entails an extraction 
from $L$ of a SK $K=g(L)$ of entropy rate $\frac{1}{n}H(L|{\bf F})$, by 
means of a suitable operation performed {\it identically} at each terminal 
on the acquired common randomness $L$. When the common randomness first 
acquired by the $d$ terminals is maximal, i.e., $L=({\bf X}_1, \cdots , 
{\bf X}_d)$ with probability $\cong 1$, then the corresponding SK $K=g(L)$ 
has the best rate $C_{SK}$ given by (\ref{e1}). A similar approach is used 
to generate a PK of rate given by (\ref{e2}).

The discussion above suggests that techniques for multiterminal SW data 
compression could be used for the {\it construction} of SKs and PKs. Next, 
in SW coding, the existence of linear data compression codes with rates 
arbitrarily close to the SW bound has been long known \cite{Csi82}. 
In particular, when the i.i.d. sequences observed at the terminals are 
related to each other through virtual communication channels characterized 
by independent additive noises, such linear data compression codes can be 
obtained in terms of the cosets of linear error-correction codes for these 
virtual channels, a fact first illustrated in \cite{Wyn74} for the special 
case of $d=2$ terminals connected by a virtual binary symmetric channel (BSC). 
This fact, exploited by most known linear constructions of SW codes 
(cf. e.g. \cite{ColLee04, GarZha01, LivXio02, PraRam03}), can enable us to 
translate these constructions and other significant recent developments in 
capacity-achieving linear codes into new SK and PK constructions. (See also 
recent independent work \cite{Mur04} for related existence results, as also 
\cite{ThaDih04}.)

Motivated by these considerations, we seek to devise new 
{\it constructive schemes} for secrecy generation. The main technical 
contribution of this work is the following: we consider four simple models 
of secrecy generation and show how a new class of secret and private keys 
can be constructed, based on the SW data compression code from \cite{Wyn74}. 
While we do not specify exactly the linear capacity-achieving channel codes 
used in the SW step of the procedure, these can be chosen -- for instance -- 
from the class of LDPC \cite{LivXio02} and turbo codes \cite{GarZha01} that 
have attracted wide attention.

\section{Preliminaries}
Consider a DMMS with $d\geq 2$ components, with corresponding generic 
rvs $X_1,\cdots ,X_d$ taking values in finite alphabets 
${\cal X}_1, \cdots ,{\cal X}_d$, respectively. Let 
${\bf X}_i=(X_{i,1}, \cdots , X_{i,n})$, $i\in {\cal M}=\{1, \cdots , d\}$, 
be $n$ i.i.d. repetitions of rv $X_i$. Terminals $1, \cdots ,d$, with 
respective observations ${\bf X}_1, \cdots , {\bf X}_d$, represent the 
$d$ users who wish to generate a SK by public communication. These 
terminals can communicate with each other through broadcasts over a 
noiseless public channel, possibly interactively in many rounds. 
In general, a transmission from a terminal is allowed to be any function 
of its observations, and of all previous transmissions.  Let ${\bf F}$ 
denote collectively all the public transmissions.

Given $\varepsilon>0$, the rv $K_{\cal S}$ represents an 
{\it $\varepsilon$-secret key} ({\it $\varepsilon$-SK}) for the terminals 
in ${\cal M}$, achieved with communication ${\bf F}$, if there exist rvs 
$K_i=K_i({\bf X}_i, {\bf F})$, $i\in {\cal M}$, with $K_i$ and $K_{\cal S}$ 
taking values in the same finite set ${\cal K_S}$ such that $K_{\cal S}$ satisfies 

$\bullet$ the common randomness condition
\[
\Pr( K_i=K_{\cal S},\ i\in {\cal M})\geq 1-\varepsilon;
\]

$\bullet$ the secrecy condition
\[
\frac{1}{n}I(K_{\cal S}\wedge {\bf F})\leq \varepsilon;
\]

$\bullet$ the uniformity condition
\[
\frac{1}{n}H(K_{\cal S})\geq \frac{1}{n}\log |{\cal K}_{\cal S}|-\varepsilon.
\]

Let $A\subset {\cal M}$ be an arbitrary subset of terminals. The rv 
$K_{\cal P}(A)$ represents an {\it $\varepsilon$-private key} 
({\it $\varepsilon$-PK}) for the terminals in $A$, private from the 
terminals in $A^c={\cal M}\backslash A$, achieved with communication 
${\bf F}$, if there  exist rvs $K_i=K_i({\bf X}_i, {\bf F})$, $i\in A$, 
with $K_i$ and $K_{\cal P}(A)$ taking values in the same finite set 
${\cal K_P}(A)$ such that $K_{\cal P}(A)$ satisfies 

$\bullet$ the common randomness condition
\[
\Pr(K_i=K_{\cal P}(A), i\in A)\geq 1-\varepsilon;
\]

$\bullet$ the secrecy condition
\[
\frac{1}{n}I\left(K_{\cal P}(A)\wedge \{{\bf X}_{i},\  i\in A^c\}, {\bf F}\right)\leq \varepsilon;
\]

$\bullet$ the uniformity condition
\[
\frac{1}{n}H(K_{\cal P}(A))\geq \frac{1}{n}\log \left|{\cal K}_{\cal P}(A)\right|-\varepsilon.
\]

{\bf Definition 1} \cite{CsiNar04}: A nonnegative number $R$ is called 
an {\it achievable SK rate} if an $\varepsilon_n$-SK $K_{\cal S}^{(n)}$ 
is achievable with suitable communication (with the number of rounds 
possibly depending on $n$), such that $\varepsilon_n\rightarrow 0$ and 
$\frac{1}{n}H\left(K_{\cal S}^{(n)}\right)\rightarrow R$. The largest 
achievable SK rate is called the {\it SK-capacity}, denoted by $C_{SK}$. 
The PK-capacity for the terminals in $A$, denoted by $C_{PK}(A)$, is 
similarly defined. An achievable SK rate (resp. PK rate) will be called 
strongly achievable if $\varepsilon_n$ above can be taken to vanish 
exponentially in $n$. The corresponding capacities are termed strong capacities. 

Single-letter characterizations have been provided for $C_{SK}$ in the 
case of $d=2$ terminals in \cite{Mau93, AhlCsi93} and for $d\geq 2$ in 
\cite{CsiNar04}; and for $C_{PK}(A)$ in case of $d=3$ in \cite{AhlCsi93} 
and for $d\geq 3$ in \cite{CsiNar04}. The proofs of the achievability 
parts exploit the close connection between secrecy generation and SW data 
compression. For instance, ``common randomness,'' without any secrecy 
restrictions, is first generated through SW-compressed interterminal
communication. This means that all the $d$ terminals
acquire a rv with probability $\cong 1$. In the next step, secrecy is 
then extracted from this common randomness by means of a suitable 
{\em identical} operation performed at each terminal 
on the acquired common randomness. When the common randomness
first acquired by the $d$ terminals is maximal, then the corresponding 
secret key has the best
rate $C_{SK}$ given by (\ref{e1}). 

In this work, we consider four simple models for which we illustrate 
the {\it construction} of appropriate {\it strong} secret or private 
keys, which rely on suitable SW codes. The SW codes of interest will 
rely on the following result concerning the existence of ``good'' 
linear channel codes for a BSC. 

Hereafter, a BSC with crossover probability $p$, $0< p< \frac{1}{2}$, 
will be denoted by BSC($p$). Let $h_b(p)$ be the binary entropy function.

{\bf Lemma 1} \cite{Eli55}: For each $\varepsilon>0$, $0< p < \frac{1}{2}$, 
and for all $n$ sufficiently large, there exists a binary linear $(n,n-m)$ 
code for the BSC($p$), where $m<n[h_b(p)+\varepsilon]$, such that the average 
error probability of maximum likelihood decoding is less than $2^{-n \eta}$, 
for some $\eta>0$.

\section{Main Results}

\noindent{\bf MODEL 1:} {\it Let the terminals $1$ and $2$ observe, respectively,
$n$ i.i.d. repetitions of the correlated rvs $X_1$ and $X_2$, where
$X_1$, $X_2$ are $\{0,1\}$-valued rvs with joint probability mass function (pmf)
  \begin{equation}\label{bsc}
P_{X_1X_2}(x_1,x_2)=\frac{1}{2}(1-p)\delta_{x_1x_2}+\frac{1}{2}p\ 
(1-\delta_{x_1x_2}), \ \ p<\frac{1}{2},
  \end{equation}
with $\delta$ being the Kronecker delta function. These two terminals 
wish to generate a strong SK of maximal rate.}

The SK-capacity for this model is \cite{Mau93, AhlCsi93, CsiNar04}
\[
C_{SK}=I(X_1\wedge X_2)=1-h_b(p)\  bit/symbol.
\]
In the following, we show a simple scheme for both terminals to 
generate a SK with rate close to $1-h_b(p)$, which relies on Wyner's 
well-known method for SW data compression \cite{Wyn74}. The SW problem 
of interest entails terminal $2$ reconstructing the observed
sequence ${\bf x}_1$ at terminal $1$ from the SW codeword for ${\bf x}_1$
and its own observed sequence ${\bf x}_2$.

\noindent {\it (i) SW data compression} \cite{Wyn74}: Let ${\cal C}$ be 
the linear $(n,n-m)$ code specified in Lemma 1 with parity check matrix 
${\bf P}$. Both terminals know ${\cal C}$ and ${\bf P}$. 

Terminal $1$ transmits the syndrome ${\bf P}{\bf x}_1^t$ to terminal $2$. 
The maximum likelihood estimate of ${\bf x}_1$ at terminal 2 is: 
\[
{\hat {\bf x}_2}(1)={\bf x}_2\oplus f_{\bf P}({\bf P}{\bf x}_1^t\oplus 
{\bf P}{\bf x}_2^t),
\]
where $f_{\bf P}({\bf P}{\bf x}_1^t\oplus {\bf P}{\bf x}_2^t)$ is the 
most likely $n$-sequence ${\bf v}$ with syndrome 
${\bf P}{\bf v}^t={\bf P}{\bf x}_1^t\oplus {\bf P}{\bf x}_2^t$, with 
$\oplus$ denoting addition modulo 2 and $t$ denoting transposition.

The probability of decoding error at terminal $2$ is given by 
\[
\Pr({\hat {\bf X}_2}(1)\neq {\bf X}_1) = \Pr ({\bf X}_2\oplus f_{\bf P}
({\bf P}{\bf X}_1^t\oplus {\bf P}{\bf X}_2^t)\neq {\bf X}_1).
\]
Under the given joint pmf (\ref{bsc}), ${\bf X}_2$ can be considered as 
an input to a virtual BSC($p$), while ${\bf X}_1$ is the corresponding 
output, i.e., we can write 
\[
{\bf X}_1={\bf X}_2\oplus {\bf V},
\]
where ${\bf V}=(V_1,\cdots ,V_n)$ is an i.i.d. sequence of 
$\{0, 1\}$-valued rvs, independent of ${\bf X}_2$, with $\Pr(V_i=1)=p$, 
$1\leq i\leq n$. It readily follows that
\[
\Pr({\hat {\bf X}_2}(1)\neq {\bf X}_1) = \Pr(f_{\bf P}({\bf P}{\bf V}^t)\neq {\bf V}).
\]
Therefore, it follows from Lemma 1 that for some $\eta>0$, 
\[
\Pr({\hat {\bf X}_2}(1)\neq {\bf X}_1)<2^{-n \eta},
\]
for all $n$ sufficiently large.

\noindent {\it (ii) SK construction}: Consider a (common) standard 
array for ${\cal C}$ known to both terminals. Denote by ${\bf a}_{i,j}$ 
the element of the $i^{th}$ row and the $j^{th}$ column in the standard 
array, $1\leq i\leq 2^m$, $1\leq j\leq 2^{n-m}$. 

Terminal $1$ sets $K_{1}=j_1$ if ${\bf X}_1$ equals ${\bf a}_{i,j_1}$ in 
the standard array. Terminal $2$ sets $K_{2}=j_2$ if ${\hat {\bf X}_2}(1)$ 
equals ${\bf a}_{i,j_2}$ in the same standard array.

\noindent {\it (iii) SK criteria}: The following theorem shows that $K_1$ 
constitutes a strongly achievable SK with rate approaching the SK-capacity.

{\bf Theorem 1}: The pair of rvs $(K_1, K_2)$ generated above, with (common) 
range ${\cal K}_1$ (say), satisfy 
\[
\Pr(K_{1}\neq K_{2}) <2^{-n \eta};
\]
\[
I(K_1\wedge {\bf F})=0;
\]
\[
H(K_1)=\log |{\cal K}_1|.
\]
Further, 
\[
\frac{1}{n}H(K_1)> 1-h_b(p)-\varepsilon.
\]

{\it Remark}: The probability of $K_{1}$ being different from $K_{2}$ 
exactly equals the average error probability of maximum likelihood decoding 
when ${\cal C}$ is used on a BSC($p$). Furthermore, the gap between the rate 
of the generated SK and the SK-capacity is as wide as the gap between the 
rate of ${\cal C}$ and the channel capacity. Therefore, if a ``better'' 
channel code for a BSC($p$), in the sense that the rate of this code is closer 
to the channel capacity and the average error probability of maximum likelihood 
decoding is smaller, is applied, then a ``better'' SK can be generated at both 
terminals, in the sense that the rate of this SK is closer to the SK-capacity 
and the probability is smaller that the keys generated at different terminals 
do not agree with each other.

\noindent{\bf MODEL 2:} {\it Let the terminals $1$ and $2$ observe, respectively,
$n$ i.i.d. repetitions 
of the correlated rvs $X_1$ and $X_2$, where
$X_1$, $X_2$ are $\{0,1\}$-valued rvs with
joint pmf
\begin{eqnarray}
P_{X_1X_2}(0,0)&=&(1-p)(1-q),\nonumber\\
P_{X_1X_2}(0,1)&=&pq, \nonumber\\
P_{X_1X_2}(1,0)&=&p(1-q),\nonumber\\
P_{X_1X_2}(1,1)&=&q(1-p),\nonumber
\end{eqnarray}
where $p< \frac{1}{2}$ and $0< q< 1$. These two terminals wish to generate 
a strong SK of maximal rate.}

Note that Model 1 is a special case of Model 2 for $q=\frac{1}{2}$. We show 
below a scheme for both terminals to generate a SK with rate close to the 
SK-capacity for this model \cite{Mau93, AhlCsi93, CsiNar04}, which is 
\[
C_{SK}=I(X_1\wedge X_2)= h_b(p+q-2pq)-h_b(p)\ bit/symbol.
\]

\noindent {\it (i) SW data compression}: This step is identical to step 
{\it (i)} for Model 1 .

\noindent {\it (ii) SK construction}: Suppose that both terminals know the 
linear $(n,n-m)$ code ${\cal C}$ specified in Lemma 1, and a (common) 
standard array for ${\cal C}$. Let $\{ {\bf e}_i:1\leq i\leq 2^m\}$ denote 
the set of coset leaders for all the cosets of ${\cal C}$. Given a (generic) 
$\{0,1\}$-valued rv $X$, the set of sequences ${\bf x}\in \{0,1\}^n$ is 
called {\it $X$-typical with constant $\xi$}, denoted by $T_{X,\xi}^n$, if
\[
2^{-n[H(X)+\xi]}\leq P_{X}^n({\bf x})\leq 2^{-n[H(X)-\xi]}.
%\label{tp1}
\]
Denote by $A_i$ the set of $T_{X_1,\xi}^n$-sequences in the coset of ${\cal C}$ 
with coset leader ${\bf e}_i$, $1\leq i\leq 2^m$. If the number of sequences of 
the same type (cf. \cite{CsiKor81}) in $A_i$ is more than 
$2^{n[I(X_1\wedge X_2)-\varepsilon']}$, where $\varepsilon'>\xi+\varepsilon$, 
then collect arbitrarily $2^{n[I(X_1\wedge X_2)-\varepsilon']}$ such sequences 
to compose a subset, which we call a {\it regular subset} (as it consists of 
sequences of the same type). Continue this procedure until the number of 
sequences of every type in $A_i$ is less than $2^{n[I(X_1\wedge X_2)-\varepsilon']}$. 
Let $N_i$ denote the number of distinct regular subsets of $A_i$.

Enumerate (in any way) the sequences in each regular subset. Let ${\bf b}_{i,j,k}$, 
where $1\leq i\leq 2^m$, $1\leq j\leq N_i$, $1\leq k\leq 2^{n[I(X_1\wedge X_2)-\varepsilon']}$, 
denote the $k^{th}$ sequence of the $j^{th}$ regular subset in the $i^{th}$ coset (i.e., 
the coset with coset leader ${\bf e}_i$).

Terminal $1$ sets $K_{1}=k_1$ if ${\bf X}_1$ equals ${\bf b}_{i,j_1,k_1}$. Otherwise, 
$K_{1}$ is set to be uniformly distributed on 
$\left\{1, \cdots, 2^{n[I(X_1\wedge X_2)-\varepsilon']}\right\}$, and independent of 
$({\bf X}_1, {\bf X}_2)$. Terminal $2$ sets $K_{2}=k_2$ if ${\hat {\bf X}_2}(1)$ 
equals ${\bf b}_{i,j_2,k_2}$. Otherwise, $K_{2}$ is set to be uniformly distributed 
on $\left\{1, \cdots, 2^{n[I(X_1\wedge X_2)-\varepsilon']}\right\}$, independent of 
$({\bf X}_1, {\bf X}_2, K_{1})$.

\noindent {\it (iii) SK criteria}: The following theorem shows that $K_1$ 
constitutes a strongly achievable SK with rate approaching the SK-capacity.

{\bf Theorem 2}: For some $\eta'=\eta'(\eta, \xi, \varepsilon, \varepsilon')>0$, 
the pair of rvs $(K_1, K_2)$ generated above, with range ${\cal K}_1$ (say), satisfy 
\[
\Pr(K_{1}\neq K_{2}) <2^{-n \eta'};
\]
\[
I(K_1\wedge {\bf F})=0;
\]
\[
H(K_1)=\log |{\cal K}_1|.
\]
Further,
\[
\frac{1}{n}H(K_1)= I(X_1\wedge X_2)-\varepsilon'.
\]

\noindent{\bf MODEL 3:} {\it Let the terminals $1,\cdots ,d$ observe, respectively,
$n$ i.i.d. repetitions of $\{0,1\}$-valued rvs $X_1, \cdots ,X_d$ which form a Markov chain 
\[
X_1 -\!\!\circ\!\!- X_2 -\!\!\circ\!\!- \cdots -\!\!\circ\!\!-   X_d,
\] 
with a joint pmf $P_{X_1\cdots X_d}$ given by: for $1\leq i\leq d-1$, 
\[
P_{X_iX_{i+1}}(x_i,x_{i+1})=\frac{1}{2}(1-p_i)\delta_{x_ix_{i+1}}
+\frac{1}{2}p_i\ (1-\delta_{x_ix_{i+1}}), \ \ p_i<\frac{1}{2}.
%\label{bsc2}
\]
These $d$ terminals wish to generate a strong SK of maximal rate.} 

Note that Model 1 is a special case of Model 3 for $d=2$. Without any 
loss of generality, let 
\[
p_j=\max_{1\leq i\leq d-1} p_i .
\]
Then, the SK-capacity for this model is \cite{CsiNar04}
\[
C_{SK}=I(X_j\wedge X_{j+1})=1-h_b(p_j)\ bit/symbol.
\]
We show below how to extract a SK with rate close to $1-h_b(p_j)$ by using 
a SW data compression scheme for reconstructing ${\bf x}_j$ at all the terminals.

\noindent {\it (i) SW data compression}: Let ${\cal C}$ be the linear 
$(n,n-m)$ code specified in Lemma 1 for the BSC($p_j$), with parity check 
matrix ${\bf P}$. Terminals $i$, $1\leq i\leq d-1$, transmit the syndromes 
${\bf P}{\bf x}_i^t$, respectively.

Let ${\hat {\bf x}}_{i}(j)$ denote the maximum likelihood estimate at 
terminal $i$ of ${\bf x}_j$. For $1\leq i\leq j-1$, terminal $i$, with the 
knowledge of (${\bf P}{\bf x}_{i+1}^t, \cdots , {\bf P}{\bf x}_{j}^t$, ${\bf x}_i$), 
forms the following successive maximum likelihood estimates
\begin{eqnarray*}
{\hat {\bf x}}_{i}(i+1)&=&{\bf x}_i\oplus f_{\bf P}({\bf P}{\bf x}_i^t\oplus {\bf P}{\bf x}_{i+1}^t),\\
{\hat {\bf x}}_{i}(i+2)&=&{\hat {\bf x}}_i(i+1)\oplus f_{\bf P}({\bf P}{\bf x}_{i+1}^t\oplus {\bf P}{\bf x}_{i+2}^t),\\
&\vdots &\\
{\hat {\bf x}}_{i}(j)&=&{\hat {\bf x}}_i(j-1)\oplus f_{\bf P}({\bf P}{\bf x}_{j-1}^t\oplus {\bf P}{\bf x}_{j}^t).
\end{eqnarray*}
For $j+1\leq i\leq d$, terminal $i$, with the knowledge of (${\bf P}{\bf x}_{j}^t, \cdots , {\bf P}{\bf x}_{i-1}^t, {\bf x}_i$), forms the following successive maximum likelihood estimates
\begin{eqnarray*}
{\hat {\bf x}}_{i}(i-1)&=&{\bf x}_i\oplus f_{\bf P}({\bf P}{\bf x}_i^t\oplus {\bf P}{\bf x}_{i-1}^t),\\
{\hat {\bf x}}_{i}(i-2)&=&{\hat {\bf x}}_i(i-1)\oplus f_{\bf P}({\bf P}{\bf x}_{i-1}^t\oplus {\bf P}{\bf x}_{i-2}^t),\\
&\vdots &\\
{\hat {\bf x}}_{i}(j)&=&{\hat {\bf x}}_i(j+1)\oplus f_{\bf P}({\bf P}{\bf x}_{j+1}^t\oplus {\bf P}{\bf x}_{j}^t).
\end{eqnarray*}
It can be shown that for some $\eta'=\eta'(\eta, d)>0$,
\[
\Pr ({\hat {\bf X}}_{i}(j)= {\bf X}_j, 1\leq i\neq j\leq d )> 1-2^{-n \eta'}.
\]

\noindent {\it (ii) SK construction}: Consider a (common) standard 
array for ${\cal C}$ known to all the terminals. Denote by ${\bf a}_{l,k}$ 
the element of the $l^{th}$ row and the $k^{th}$ column in the standard 
array, $1\leq l\leq 2^m$, $1\leq k\leq 2^{n-m}$. 

Terminal $j$ sets $K_{j}=k_j$ if ${\bf X}_j$ equals ${\bf a}_{l,k_j}$ in 
the standard array. Terminal $i$, $1\leq i\neq j\leq d$, sets $K_{i}=k_i$ 
if ${\hat {\bf X}_i}(j)$ equals ${\bf a}_{l,k_i}$ in the same standard array.

\noindent {\it (iii) SK criteria}: The following theorem shows that $K_j$ 
constitutes a strongly achievable SK with rate approaching the SK-capacity.

{\bf Theorem 3}: The set of rvs $(K_1,\cdots , K_d)$ generated above, 
with range ${\cal K}_j$ (say), satisfy 
\[
\Pr(K_{1}=\cdots = K_{d}) > 1- 2^{-n \eta'};
\]
\[
I(K_j\wedge {\bf F})=0;
\]
\[
H(K_j)=\log |{\cal K}_j|.
\]
Further,
\[
\frac{1}{n}H(K_j)> 1-h_b(p_j)-\varepsilon.
\]

\noindent{\bf MODEL 4:} {\it Let the terminals 1, 2 and 3 observe, respectively,
$n$ i.i.d. repetitions of the correlated rvs $X_1$, $X_2$, $X_3$, where $X_1$, 
$X_2$, $X_3$ are $\{0, 1\}$-valued rvs with joint pmf 
\begin{eqnarray}
&&P_{X_1X_2X_3}(0,0,0)=P_{X_1X_2X_3}(0,1,1)=\frac{(1-p)(1-q)}{2},\nonumber\\
&&P_{X_1X_2X_3}(0,0,1)=P_{X_1X_2X_3}(0,1,0)=\frac{pq}{2}, \nonumber\\
&&P_{X_1X_2X_3}(1,0,0)=P_{X_1X_2X_3}(1,1,1)=\frac{p(1-q)}{2},\nonumber\\
&&P_{X_1X_2X_3}(1,0,1)=P_{X_1X_2X_3}(1,1,0)=\frac{q(1-p)}{2},\nonumber
\end{eqnarray}
where $p< \frac{1}{2}$ and $0< q< 1$. Terminals 1 and 2 wish to generate a 
strong PK of maximal rate, which is concealed from the helper terminal 3.}

Note that under the given joint pmf of $X_1$, $X_2$, $X_3$, we can write 
\[
{\bf X}_1={\bf X}_2\oplus {\bf X}_3 \oplus {\bf V},
\]
where ${\bf V}=(V_1, \cdots ,V_n)$ is an i.i.d. sequence of $\{0, 1\}$-valued 
rvs, independent of $({\bf X}_2, {\bf X}_3)$, with $\Pr(V_i=1)=p$, $1\leq i\leq n$.

We show below a scheme for terminals 1 and 2 to generate a PK with rate close 
to the PK-capacity for this model \cite{AhlCsi93, CsiNar04}
\begin{eqnarray*}
C_{PK}(\{1,2\})&=&I(X_1\wedge X_2|X_3)\\
               &=& h_b(p+q-2pq)-h_b(p)\ bit/symbol.
\end{eqnarray*}
The preliminary step of this scheme entails terminal 3 simply revealing its 
observations ${\bf x}_3$ to both terminals 1 and 2. Then, Wyner's SW data 
compression scheme is used for reconstructing ${\bf x}_1$ at terminal 2 
from the SW codeword for ${\bf x}_1$ and ${\bf x}_2\oplus {\bf x}_3$.

\noindent {\it (i) SW data compression}: This step is identical to step 
{\it (i)} for Model 1.

\noindent {\it (ii) PK construction}: Suppose that terminals 1 and 2 know 
the linear $(n,n-m)$ code ${\cal C}$ specified in Lemma 1, and a (common) 
standard array for ${\cal C}$. Let $\{ {\bf e}_i:1\leq i\leq 2^m\}$ denote 
the set of coset leaders for all the cosets of ${\cal C}$. Given (generic) 
$\{0,1\}$-valued rvs $X$, $Y$, the set of pairs of sequences 
$({\bf x}, {\bf y})\in \{0,1\}^n\times \{0,1\}^n$ is called 
{\it $XY$-typical with constant $\xi$}, denoted by $T_{XY,\xi}^n$, if 
${\bf x}\in T_{X,\xi}^n$, ${\bf y}\in T_{Y, \xi}^n$, and 
\[
2^{-n[H(X, Y)+\xi]}\leq P_{XY}^n({\bf x}, {\bf y})\leq 2^{-n[H(X, Y)-\xi]}.
%\label{tp2}
\]
For every ${\bf y}\in \{0,1\}^n$, the set of sequences ${\bf x}\in \{0,1\}^n$ 
is called {\it $X|Y$-typical with respect to ${\bf y}$ with constant $\xi$}, 
denoted by $T_{X|Y,\xi}^n({\bf y})$, if $({\bf x}, {\bf y})\in T_{XY,\xi}^n$. 
Note that $T_{X|Y, \xi}^n({\bf y})$ is an empty set if ${\bf y}\not\in T_{Y,\xi}^n$.

For a sequence ${\bf x}_3\in \{0,1\}^n$, denote by $A_i({\bf x}_3)$ the 
set of $T_{X_1|X_3,\xi}^n({\bf x}_3)$-sequences in the coset of ${\cal C}$ 
with coset leader ${\bf e}_i$, $1\leq i\leq 2^m$. If the number of sequences 
of the same joint type (cf. \cite{CsiKor81}) with ${\bf x}_3$ in $A_i({\bf x}_3)$ 
is more than $2^{n[I(X_1\wedge X_2|X_3)-\varepsilon']}$, where 
$\varepsilon'>2\xi+\varepsilon$, then collect arbitrarily 
$2^{n[I(X_1\wedge X_2|X_3)-\varepsilon']}$ such sequences to compose a regular 
subset. Continue this procedure until the number of sequences of every joint type 
with ${\bf x}_3$ in $A_i({\bf x}_3)$ is less than 
$2^{n[I(X_1\wedge X_2|X_3)-\varepsilon']}$. Let $N_i({\bf x}_3)$ denote the 
number of distinct regular subsets of $A_i({\bf x}_3)$. 

For a given sequence ${\bf x}_3$, enumerate (in any way) the sequences 
in each regular subset. Let ${\bf b}_{i,j,k}({\bf x}_3)$, where 
$1\leq i\leq 2^m$, $1\leq j\leq N_i({\bf x}_3)$, 
$1\leq k\leq 2^{n[I(X_1\wedge X_2|X_3)-\varepsilon']}$, denote the $k^{th}$ 
sequence of the $j^{th}$ regular subset in the $i^{th}$ coset.

Terminal $1$ sets $K_{1}=k_1$ if ${\bf X}_1$ equals ${\bf b}_{i,j_1,k_1}({\bf X}_3)$. 
Otherwise, $K_{1}$ is set to be uniformly distributed on 
$\{1, \cdots, 2^{n[I(X_1\wedge X_2|X_3)-\varepsilon']}\}$, independent of 
$({\bf X}_1, {\bf X}_2,{\bf X}_3)$. Terminal 2 sets $K_{2}=k_2$ if 
${\hat {\bf X}_2}(1)$ equals ${\bf b}_{i,j_2,k_2}({\bf X}_3)$. Otherwise, $K_{2}$ 
is set to be uniformly distributed on 
$\{1, \cdots, 2^{n[I(X_1\wedge X_2|X_3)-\varepsilon']}\}$, independent of 
$({\bf X}_1, {\bf X}_2,{\bf X}_3, K_{1})$.

\noindent {\it (iii) SK criteria}: The following theorem shows that $K_1$ 
constitutes a strongly achievable PK with rate approaching the PK-capacity.

{\bf Theorem 4}: For some $\eta'=\eta'(\eta, \xi, \varepsilon, \varepsilon')>0$, 
the pair of rvs $(K_1, K_2)$ generated above, with range ${\cal K}_1$ (say), satisfy 
\[
\Pr(K_{1}\neq K_{2}) <2^{-n \eta'};
\]
\[
I(K_1\wedge {\bf X}_3, {\bf F})=0;
\]
\[
H(K_1)=\log |{\cal K}_1|.
\]
Further,
\[
\frac{1}{n}H(K_1)= I(X_1\wedge X_2|X_3)-\varepsilon'.
\]


\begin{thebibliography}{5}

\bibitem{AhlCsi93}
R.~Ahlswede and I.~Csisz\'ar, ``Common randomness in
information theory and cryptography, Part I: Secret
sharing,'' {\it IEEE Trans. Inform. Theory}, 
vol.~39, pp.~1121--1132, July 1993.

\bibitem{ColLee04}
T. P. Coleman, A. H. Lee, M. M{\'e}dard, and M. Effros, ``On some new 
approaches to practical Slepian-Wolf compression inspired by channel coding,''
{\it Proc. IEEE Data Compression Conference,\/} pp. 282--291, Snowbird, UT, March 2004.

\bibitem{Csi82}
I.~Csisz\'ar, ``Linear codes for sources and source networks: Error exponents, 
universal coding,'' {\it IEEE Trans. Inform. Theory,} vol. 28, no. 4, pp. 585--592, July, 1982.

\bibitem{CsiKor81} 
I. Csisz\'ar and J. K\"{o}rner, {\sl Information Theory: Coding Theorems for 
Discrete Memoryless Systems.\/} Academic, New York, N.Y., 1982.

\bibitem{CsiNar04}
I.~Csisz\'ar and P. Narayan, ``Secrecy capacities for multiple terminals,'' 
{\it IEEE Trans. Inform. Theory}, vol. 50, pp. 3047--3061, Dec. 2004.

\bibitem{Eli55}
P. Elias, ``Coding for noisy channels,'' {\it IRE Convention Record}, 
Part 4, pp. 37--46, 1955.

\bibitem{GarZha01}
J. Garcia-Frias and Y Zhao, ``Compression of correlated binary sources using 
turbo codes,'' {\it IEEE Commun. Lett.,} vol. 5, pp. 417--419, Oct. 2001.

\bibitem{LivXio02}
A. D. Liveris, Z. Xiong, C. N. Georghiades, ``Compression of binary sources 
with side information at the decoding using LDPC codes,'' 
{\it IEEE Commun. Lett.}, vol. 6, pp. 440--442, Oct. 2002.

\bibitem{Mau93}
U.~M.~Maurer, ``Secret key agreement by public discussion
from common information,'' {\it IEEE Trans. Inform. Theory}, 
vol.~39, pp.~733--742, May 1993.

\bibitem{Mur04}
J. Muramatsu, ``Secret key agreement from correlated source outputs using 
LDPC matrices,'' {\it IEICE Trans. Fundamentals,} vol. E87-A, 2004.

\bibitem{PraRam03}
S. S. Pradhan and K. Ramchandran, ``Distributed source coding using syndromes 
(DISCUS): Design and construction,'' {\it IEEE Trans. Inform. Theory,\/} 
vol. 49, pp. 626--643, March 2003. 

\bibitem{ThaDih04}
A. Thangaraj, S. Dihidar, A. R. Calderbank, S. McLaughlin and J. M. Merolla, 
``Capacity achieving codes for the wiretap channel with applications to 
quantum key distribution,'' e-print cs. IT/0411003, 2004.

\bibitem{Wyn74}
A. D. Wyner, ``Recent results in the Shannon theory,'' {\it IEEE Trans. 
Inform. Theory,\/} vol. 20, pp. 2--10, Jan. 1974. 

\end{thebibliography}
\end{document}